\documentclass[twocolumn,showpacs,preprintnumbers,amsmath,amssymb,prl]{revtex4}
\usepackage{dcolumn}% Align table columns on decimal point
\usepackage{bm}% bold math
 
\usepackage{amsmath}
\usepackage{amsfonts}    
\usepackage{graphics}  
\usepackage{psfrag}  
\usepackage{hyperref}
\usepackage{graphicx} 
\usepackage{epsfig}    
\usepackage{rotating}  
 \usepackage[latin1]{inputenc}
 \usepackage{color}
 \usepackage{rotating}
\begin{document}

\preprint{LMU-ASC 66/07}

\title{Noise and Correlations in a Spatial Population 
Model with Cyclic Competition}
% Force line breaks with \\  

\author{Tobias Reichenbach, Mauro Mobilia$^*$, and Erwin Frey}
\affiliation{Arnold Sommerfeld Center for Theoretical Physics (ASC) and
  Center for NanoScience (CeNS), Department of Physics,
  Ludwig-Maximilians-Universit\"at M\"unchen, Theresienstrasse 37,
  D-80333 M\"unchen, Germany}

  \begin{abstract}
Noise and spatial degrees of freedom characterize most ecosystems. 
Some aspects of their influence on the coevolution 
of populations with cyclic interspecies competition have been demonstrated in recent
experiments [e.g. B. Kerr et al.,  Nature {\bf 418}, 171 (2002)]. To reach a better
theoretical understanding
of these phenomena, we consider a  paradigmatic spatial model where three species
exhibit cyclic dominance. Using an individual-based description, as well as
stochastic 
partial differential and deterministic reaction-diffusion equations, we
account for stochastic fluctuations and spatial diffusion at  different levels, and
show how fascinating patterns of entangled spirals emerge.
We rationalize our analysis by computing the spatio-temporal correlation functions 
and provide analytical expressions for
the front velocity and the wavelength of the propagating spiral waves.
\end{abstract}

\maketitle

Understanding the combined influence 
of spatial degrees of freedom and noise on biodiversity is an important issue in
theoretical biology and
ecology. This implies to face the challenging problem of studying complex
nonequilibrium
structures, which form in the course of nonlinear
evolution~\cite{turing-1952-237,May,Murray,kerr-2002-418,tainaka-1989-63, reichenbach-2007-448}. 
More generally, self-organized nonequilibrium patterns and traveling waves are
ubiquitous 
in  nature and appear, for instance, in chemical reactions, biological
systems, as well as in epidemic outbreaks~~\cite{Kapral}.
Among the most studied types of patterns are spiral waves, which are relevant to
autocatalytic chemical reactions, aggregating 
slime-mold cells and cardiac muscle tissue~\cite{zaikin-1970-225}. 
In all these {\it nonequilibrium} and {\it nonlinear} processes, as well as
in population dynamics models~\cite{turing-1952-237,Murray,tainaka-1989-63}, pattern formation  
is driven by diffusion which, together with internal noise, act as mechanisms
allowing for stabilization and coevolution of the reactants.
In this work, we consider a  paradigmatic spatially-extended $3$ species population
system with cyclic competition, which can be regarded as a simple food-chain
model~\cite{Hofbauer}. 
In fact, such a system  is inspired by recent experiments on the
coevolution of $3$ species
of bacteria in cyclic competition~\cite{kerr-2002-418}. 
Using methods of statistical physics,
we study the influence of spatial degrees of freedom and internal noise on the
coevolution of the species and on the emerging spiral patterns. In particular, we
compute the correlation functions and provide analytical expressions for the
spreading speed and  wavelength of the propagating fronts. To underpin the role of
internal noise,
the results of the stochastic description are compared  with those of the
deterministic  equations.
 
In this Letter, we investigate a stochastic spatial variant of the \emph{rock-paper-scissors game}~\cite{Hofbauer} (also referred to as cyclic Lotka-Volterra model). These kinds of systems have been studied both from a game-theoretic perspective, see e.g.~\cite{hauert-2005-73,szabo-2007-446} and references therein, and within the framework of chemical reactions~\cite{tainaka-1989-63,Frachebourg-1996-77}, revealing rich spatio-temporal behaviors (e.g. emergence of rotating spirals). While our methods have a broad range of applicability, they are illustrated for a prototypical model introduced  by May
and Leonard~\cite{may-1975-29} where $3$ species,  $A,~B$ and $C$ undergo a 
cyclic competition (codominance with rate $\sigma$) and reproduction (with rate $\mu$), according
to the reactions
\begin{eqnarray}
AB \stackrel{\sigma}{\longrightarrow} A\oslash&\,, \quad
BC\stackrel{\sigma}{\longrightarrow} B\oslash&\,, \quad  
CA \stackrel{\sigma}{\longrightarrow} C\oslash\,, \cr
A\oslash \stackrel{\mu}{\longrightarrow}AA&\,, \quad
 B\oslash\stackrel{\mu}{\longrightarrow} BB&\,, \quad 
C\oslash \stackrel{\mu}{\longrightarrow}CC\,. 
\label{ml_react}
\end{eqnarray}  
Hence, an individual of species $A$  will consume one of species $B$ ($AB\rightarrow
A\oslash$) with rate 
$\sigma$ and will reproduce with rate $\mu$ if an empty spot, denoted $\oslash$, is
available
($A\oslash\rightarrow AA$, i.e. there is a {\it finite} carrying capacity). In
addition, to mimic the possibility of migration, it is realistic to 
endow the individuals with a form of mobility. For the sake of simplicity, we consider
a simple exchange process, with rate $\epsilon$, among any {\it nearest-neighbor} 
pairs of agents:   
$XY \stackrel{\epsilon}{\longrightarrow} Y X, \text{ where \,} X, Y \in \{A,B,C, 
\oslash\}$. 
If one ignores the spatial structure and assumes the system to be well-mixed (with an
infinite number of individuals), the 
population's mobility plays no role and the dynamics is aptly described by the
deterministic rate equations (RE) for the densities
$a,b,c$ of species $A, B$ and $C$, respectively. Introducing ${\bm s}\equiv (a,b,c)$,
the RE read: 
\begin{eqnarray}
\label{RE}
\partial_t s_i= s_i[\mu(1-\rho)-\sigma s_{i+2}], \quad i\in \{1,2,3\}
\end{eqnarray}
where the index $i$ is taken modulo 3 and $\rho= a+b+c$ is the total density.
As shown by May and Leonard~\cite{may-1975-29} (see also~\cite{durrett-1998-53}),
these equations possess 4 absorbing
fixed points, corresponding to a system filled with only one species and to an empty
system. In addition,
there is a reactive fixed point ${\bm s}^*=\frac{\mu}{\sigma+3\mu}(1,1,1)$,
corresponding to a total density $\rho^*=\frac{3\mu}{\sigma+3\mu}$. A linear
stability analysis shows
that ${\bm s}^*$ is {\it unstable}. The absorbing steady states $(1,0,0)$, $(0,1,0)$
and $(0,0,1)$ are heteroclinic points. The existence of a Lyapunov function ${\cal
L}=abc/\rho^3$ 
allows to prove that, within the realm of the above RE, the phase portrait is
characterized by flows 
spiraling outward from ${\bm s^*}$, with frequency 
$\omega_0=\sqrt{3}\mu\sigma/[2(3\mu+2\sigma)]$ in its vicinity. Approaching the
boundaries of the phase portrait, the trajectories form (heteroclinic) cycles
indefinitely close to the edges (without ever reaching them), with densities
approaching in turn the value one. Despite its mathematical elegance, this behavior
has been recognized to be  unrealistic~\cite{may-1975-29,durrett-1998-53}.
 In fact, for  finite populations, fluctuations arise
and {\it always} cause the extinction of two species 
 in finite time (see e.g. Ref~\cite{reichenbach-2006-74}).   

In this work, considering the spatial version of the above model in the presence
of internal noise, we show that a robust (and, arguably, more realistic) scenario for the 
evolution arises. The reaction schemes~(\ref{ml_react}) and the exchange events are
considered to occur
 on a $d-$dimensional regular lattice of $N$ sites, labeled
${\bm r}=(r_1, \dots, r_d)$. Each lattice site has $z$  neighbors
at a distance $\delta r$ (e.g. $z=2d$ and $N=L^d$ for hypercubic lattices of linear
size $L$) and is either empty or occupied by at most one individual. 
On the lattice, the binary reactions~(\ref{ml_react}) and exchanges
only occur among pairs
of {\it nearest-neighbors}. 
In the situation of large system sizes, the continuum limit reveals that for the
exchange process to be an efficient driving mechanism, the rate $\epsilon$ has to
scale
as $\epsilon \propto N^{\nu}$, with $\nu=2/d$ and $N\to \infty$. In fact, if
$0<\nu<2/d$ the system is dominated by the local
reactions~(\ref{ml_react}) among neighboring individuals; while 
effective diffusion renders locality irrelevant when  $\nu>2/d$. Only when 
$\nu=2/d$, there is an effective competition between the stirring process and the
local reactions~(\ref{ml_react}). It is therefore useful to
introduce 
the {\it effective diffusion constant}  $D\equiv \frac{z}{2d^2} N^{-2/d}\,\epsilon$.
Because of the  discreteness of the number of individuals involved in the reactions,
internal fluctuations arise in the system. The latter originate from (i) the 
interspecies reactions~(\ref{ml_react})  and (ii) the exchange
processes. In the continuum limit, 
where $\delta r=N^{-1/d}$ with $N,\epsilon\to \infty$ (and finite $D$), there
is a separation of time scales and the pair exchanges occur much faster
than the reactions ($\epsilon \propto N^{2/d}$). Actually, the fluctuations
associated with~(\ref{ml_react}) and the agents' mobility scale
respectively as $N^{-1/2}$
and $N^{-1}$, with the former dominating over the latter and being the only relevant
contribution. This result is revealed by a system size, also called Kramers-Moyal (see e.g. Ref.~\cite{Gardiner}, Chap.~8), expansion (SZE) of the master equation underlying the  
exchange processes and the reactions~(\ref{ml_react})~\cite{EPAPS}. Furthermore, the SZE yields a proper Fokker-Planck
equation, which is equivalent to a set of (Ito) stochastic partial differential equations (SPDE) with white noise. 
The derivation, obtained in the continuum limit from the master equation, is outlined in the supplementary EPAPS document~\cite{EPAPS}
and will be detailed elsewhere \cite{reichenbach-2007-3}. Here, we quote the expression of the SPDE: 
\begin{eqnarray}
\partial_t s_i= D\nabla^2  s_i+\mathcal{A}_i\big({\bm s}\big) + \sum_{j=1}^3
\mathcal{C}_{ij}\big({\bm s}\big)\xi_j,\quad i\in \{1,2,3\}
\label{spde}
\end{eqnarray} 
where $\nabla^2$ is the Laplacian operator; $\langle \xi_i({\bm r},t)\rangle=0$, $
\langle \xi_i({\bm r},t)\xi_j({\bm r}',t')\rangle=\delta_{i,j}\delta({\bm r}-{\bm
r}')\delta(t-t')$  and
\begin{eqnarray}   
\label{A_C}
\mathcal{A}_i\big({\bm s}\big)&=&s_i[\mu(1- \rho)-\sigma s_{i+2}],\\ 
 \mathcal{C}_{ij}\big({\bm s}\big)&=&\delta_{ij}\sqrt{N^{-1}s_i\big[\mu(1-
\rho)+\sigma s_{i+2}\big]}.
\end{eqnarray} 
Again, the indices are taken modulo $3$ and now $ s_i\equiv s_i({\bm r},t)$. 
As explained in~\cite{Gardiner,reichenbach-2007-3}, these SPDE
have to be interpreted in the sense of Ito calculus.
While Eqs.~(\ref{spde}) and our approach  are valid in any dimension~\cite{EPAPS},
for specificity, we now analyze the spatio-temporal properties of the system
in two dimensions  with
periodic boundary conditions. On the one hand we have solved numerically the
SPDE~(\ref{spde})  using the open software from the XMDS project~\cite{xmds}. 
On the other hand, we have carried out individual-based simulations 
 of the reactions~(\ref{ml_react}) for mobile (exchange process)
particles  
 on lattices of size $L\times L$, with $L=30 - 1000$. This allows to carefully study
the convergence towards the continuum limit, where the description in terms of (\ref{spde}) is expected to be accurate.

\begin{figure}    
\begin{center}    
\includegraphics[scale=1]{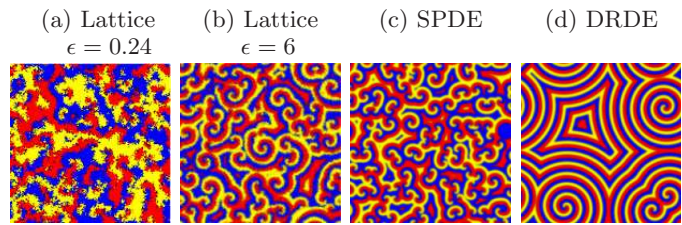} 
\caption{(Color online) Snapshots of reactive steady states for rates $D=3\times
10^{-6}, \mu=\sigma=1$. Each color (level of grey) indicates one species (black dots correspond to
vacancies). In (a) and (b)  results are from lattice simulations for $L=200$ (a) and
$L=1000$ (b), i.e. different $\epsilon$. Spiral structures emerge for sufficiently large
exchange rate (b). Numerical solution of the SPDE~(\ref{spde}) and DRDE are shown in
(c), resp. (d); see text.
In (a), (b) and (c), initially ${\bm s}({\bm r},0)={\bm s}^*$. 
\label{snap}  
} 
\end{center}                    
\end{figure}
\begin{figure}  
\begin{center}
\includegraphics[scale=1]{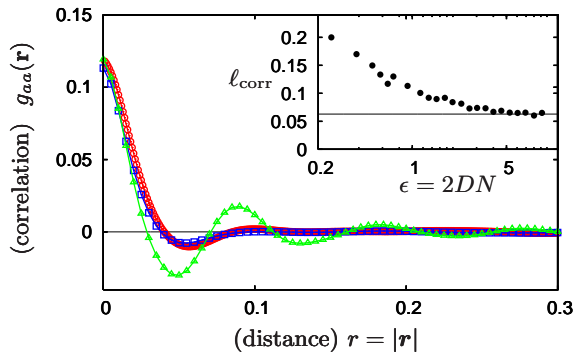}
\caption{(Color online)\quad Spatial correlation functions in 2D, obtained from 
lattice simulations (red, circles; $\epsilon=6$, $L=1000$), from the solution of the
SPDE~(\ref{spde}) (dark blue, squares) and of the DRDE (green, triangles), see text. 
The reaction rates are $\mu=\sigma=1$ and $D=3\times 10^{-6}$. Inset:
The correlation length  $\ell_\text{corr}$, for $D=5\times10^{-5}, \mu=\sigma=1$, as
function of $\epsilon$ (i.e. for different lattice sizes) compared to the prediction of the
SPDE (black line). The latter is in excellent agreement with lattice simulations 
already for $\epsilon \geq 5$ (i.e. $L\geq 225$).
\label{corr_comp}} 
\end{center}                
\end{figure}

As other spatially-extended dynamical systems~\cite{tainaka-1989-63,hauert-2005-73,szabo-2007-446,Frachebourg-1996-77}, the model under consideration
displays fascinating nonequilibrium patterns emerging in the course of the
evolution. In Fig.~\ref{snap}~(a) and (b), we report typical long-time snapshots of
the system for low (a) and high (b) exchange rates (but keeping $D$ fixed), as
obtained from lattice simulations. In both cases we notice that intriguing patterns
form. For slow exchange rate, the system displays non-geometrical patches, similarly
to what happens in systems with self-organized
criticality~\cite{schenk-2000-15}. When the exchange rate is raised, the patterns
display spiral structures. In fact, starting from a spatially homogeneous initial
condition, 
${\bm s}({\bm r},0)={\bm s}^{*}$, the system is randomly perturbed  by the internal
noise
and the resulting spatial inhomogeneities grow and form wavefronts moving through
the system. 
The emergence of spiral patterns is a feature shared by other {\it excitable systems}
(see e.g.~\cite{Kapral,zaikin-1970-225})
 and corresponds to the ability of the system to sustain the propagation of 
 oscillating waves.
For sufficiently large $\epsilon$, one observes a striking resemblance between the
size and structure of the patterns 
obtained from the lattice simulations [Fig.~\ref{snap}~(b)] and those from the
SPDE~(\ref{spde})  [Fig.~\ref{snap}~(c)]. To further compare the predictions of the SPDE~(\ref{spde}) with the lattice
simulations, 
and to gain additional information on the structure of the the emerging patterns, 
we have computed the  correlation functions, $g_{s_i  s_j}({\bm r- \bm
r}',t)\equiv\langle s_i({\bm r},t) s_j({\bm r}',t) \rangle - \langle s_i({\bm r},t)\rangle \langle s_j({\bm r}',t) \rangle$ in two dimensions. 
In Fig.~\ref{corr_comp} (red and blue curves), we report the results for $g_{aa}({\bm
r},t)$ in the steady state and notice an excellent agreement between the results of
the lattice simulations and the predictions of the SPDE~(\ref{spde}). The inset of
Fig.~\ref{corr_comp}, displays the correlation length 
$\ell_{\rm corr}$~\footnote{The correlation length denotes the characteristic length at which the spatial correlations  decay by a factor $1/e$ from their maximal value.} 
as function of $\epsilon$ ($D$ is kept fixed, $L$ varies)
obtained in the lattice simulations, which is found to coincide with the prediction
of the SPDE already for $\epsilon \geq 5$. We have also computed
the autocorrelation function $g_{s_i s_j}(0,t)$ and found, both in the
lattice simulations
and from the solutions of the SPDE, an oscillating behavior with a similar characteristic
frequency, markedly different from $\omega_0$~\cite{reichenbach-2007-3}.  
This confirms that, even for {\it finite exchange rates}, 
the solution of the SPDE~(\ref{spde}) provides an excellent approximation of the 
 lattice simulations of the system. This is rather surprising since Eqs.~(\ref{spde})
have been  derived
in the continuum limit, where $N$ and $\epsilon \to \infty$. A comparable influence
of finite exchange rate in a predator-prey system has  been
reported recently~\cite{mobilia-2006-73}.
According to the SPDE~(\ref{spde}), $\ell_{\rm corr}$ scales as $D^{1/2}$, so that by
raising
the diffusion one increases the size of the spirals. 
As we have shown in Ref.~\cite{reichenbach-2007-448}, this happens up to a critical
value $D_c$ (e.g. $D_c\approx 4.5\pm0.5 \times 10^{-4}$ for $\mu=\sigma=1$): above
that threshold, the spiral structures outgrow the system size and only one
species survives, corresponding to an absorbing steady state predicted by
Eqs.~(\ref{RE}).

As the properties of the lattice simulations are well captured by the
SPDE~(\ref{spde}), where the strength of the noise scales as $N^{-1/2}$, with $N\to
\infty$, it is natural to investigate  the  actual influence 
of this internal noise on the steady state of the system. To address this issue, we
have 
solved numerically (in 2D, with periodic boundary conditions)
the deterministic reaction-diffusion equation (DRDE) obtained from~(\ref{spde}) by
dropping the noise terms, i.e.
$ \partial_t s_i= D\nabla^2 s_i+\mathcal{A}_i\big({\bm s}\big).$
 Of course, to obtain a nontrivial steady state for the DRDE one has to assume 
spatially inhomogeneous initial conditions. In Fig.~\ref{snap} (d), we have reported
a snapshot of the 
long-time behavior predicted by 
 the DRDE starting from  ${\bm s}({\bm r},0)={\bm s}^*+ (\frac{1}{100}\cos{2\pi
r_1r_2},0,0)$. In this case, 
the dynamics evolves towards a reactive steady state which also exhibits spiral
waves. However, the latter do not form  entangled structures, but ordered geometrical
patterns. As an example, only four spirals cover the system in Fig.~\ref{snap}~(d) [while noise leads to 
106 entangled spirals in Fig.~\ref{snap}~(c)]. The correlation functions 
 associated with the DRDE
 therefore exhibit only weakly damped spatial oscillations (see Fig.~\ref{corr_comp},
green triangles).
By analyzing typical snapshots like those of Fig.~\ref{snap}~(d), we have noted that
in the deterministic
and stochastic [i.e. lattice simulations with ``large'' $\epsilon$ and solutions of
Eqs.~(\ref{spde})]  descriptions, the spiral waves
share the same  propagation velocity, frequency and wavelength.
However, a major difference between these descriptions lies on the crucial dependence
of the DRDE
on initial conditions,  which determine the overall number of spirals and their size.
On the contrary, because the internal noise acts a
random source of spatial inhomogeneities, the lattice stochastic system and the SPDE display \emph{robust features}.
In particular, we have found noise to induce a universal spiral density of about $0.5$ per square wavelength.

\begin{figure}  
\begin{center}
\includegraphics[scale=1]{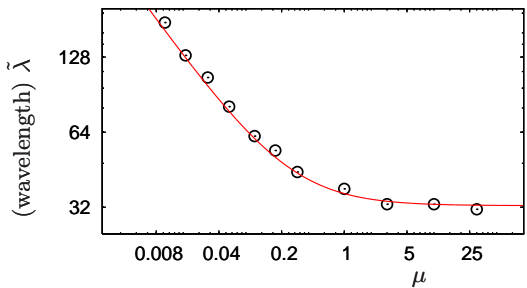}
\caption{(Color online)\quad Plot of $\tilde{\lambda}$, where
 $\lambda=\tilde{\lambda}\sqrt{D}$ is the spirals' wavelength 
  of the propagating spiral waves. Analytical results (red curve, rescaled by a
factor $1.6$; see text) are compared with the solution 
 of the SPDE~(\ref{spde}) (black circles). \label{wlength}}
\end{center}                
\end{figure}

Analytical expressions for the spreading velocity and the wavelength of the
propagating fronts of the DRDE can be 
obtained by considering the dynamics on the invariant manifold of the RE~\cite{Wiggins}, given by ${\cal M}:\{y_C=\frac{\sigma(3\mu+\sigma)}{4\mu(3\mu+ 2\sigma)}\,
(y_A^2+y_B^2) + O(y^3)\}$, with 
 $(y_A, y_B, y_C)^{T}\equiv
\frac{1}{3}
{\scriptsize
\begin{pmatrix} \sqrt{3} & 0 & -\sqrt{3} \\
                                        -1 & 2 & -1 \\
			1 & 1 & 1          \end{pmatrix}}
			 ({\bm s}^{T}-{\bm s}^{*\,T})$.
On ${\cal M}$,  up to $3^{rd}$ order, the DRDE can be recast in the form of a forced
complex Ginzburg-Landau equation (CGLE)~\cite{saarloos-2003-386,cross-1993-65}. By
performing the 
nonlinear transformation $z_A= y_A+\frac{3\mu + \sigma}{28\mu}\left[\sqrt{3}y_A^2 +
10 y_A y_B -\sqrt{3}y_B^2\right]$
and $z_B= y_B+\frac{3\mu + \sigma}{28\mu}\left[5y_A^2 + 2\sqrt{3}y_A y_B
-5y_B^2\right]$,
upon ignoring nonlinearities like $(\nabla z_{A,B})^2$, one is left with the
following CGLE in the variable $z\equiv z_A + iz_B$~\cite{reichenbach-2007-3}:
\begin{eqnarray}
\label{CGLE}
\partial_t z &=& D\nabla^2 z + (c_1-i\omega_0)z - c_2(1+ic_3)|z|^2 z ,
\end{eqnarray}
with
$c_1\equiv\frac{\mu\sigma}{2(3\mu+\sigma)}$, 
$c_2\equiv\frac{\sigma(3\mu + \sigma)(48\mu +11\sigma)}{56\mu(3\mu+2\sigma)},$ and
$c_3 \equiv\frac{\sqrt{3}(18\mu + 5\sigma)}{(48\mu +11\sigma)}$.
The general theory of front propagation~\cite{saarloos-2003-386,cross-1993-65}
predicts that Eq.~(\ref{CGLE}) always admits traveling waves as stable solutions
(i.e. no Benjamin-Feir or Eckhaus instabilities occur).
We have determined such periodic solutions by computing, from the dispersion relation
of~(\ref{CGLE}), the spreading velocity $v$
and the spirals' wavelength $\lambda$ (details will be given in \cite{reichenbach-2007-3}): 
\begin{eqnarray}
\label{res}
v= 2\sqrt{c_1 D}, \; 
\lambda=2\pi c_3
\sqrt{c_1^{-1}D}
\left(1-\sqrt{1+ c_3^2}\right)^{-1}. 
\end{eqnarray} 
In the stochastic version of the model, the wavelength and velocity of the wavefronts
have been found to agree
with those of the deterministic treatment.
Hence, the expressions~(\ref{res}) also apply (for large $\epsilon$, with $D<D_c$) to
the results of lattice simulations  (rescaled by a factor $L$) and to the solution of
the SPDE~(\ref{spde}).
For instance, on a square grid with $\mu=\sigma=1$, lattice simulations and
Eqs.~(\ref{spde}) yield
$v\approx 0.63 D^{1/2}L$, in good agreement with the prediction of~(\ref{res}):
$v=(D/2)^{1/2}L$. For the spirals' wavelength, numerical results
(lattice simulations and SPDE) yield $\lambda\propto D^{1/2}$ as predicted
by~(\ref{res}).
In Fig.~\ref{wlength}, the analytical prediction~(\ref{res}) for $\lambda$ is
compared with the 
values obtained from the SPDE~(\ref{spde}), yielding a remarkable
agreement for the functional dependence on the parameter $\mu$. Yet, as
Eq.~(\ref{CGLE}) does not account
for all nonlinearities, the analytical and numerical 
values differ by a prefactor  $\approx 1.6$ (considered in
Fig.~\ref{wlength})~\cite{reichenbach-2007-3}. It can still be noted that~(\ref{CGLE}) and the predictions~(\ref{res}) are valid in all dimensions \cite{EPAPS,reichenbach-2007-3}.

Motivated by recent experiments~\cite{kerr-2002-418}, we have considered a
spatially-extended model with three species in cyclic competition, and focused on the 
spatial and stochastic effects.
The local character of the reactions
and internal noise allow mobile populations to coexist and lead to pattern
formation.
We have shown that already for finite mobility the lattice 
model can be described by SPDE. With the latter and lattice
simulations, we have studied how entanglement of spirals form  and obtained
expressions for their spreading velocity and wavelength.  
The size of the patterns crucially depends  on the diffusivity: above a certain threshold the  system is covered by  one species~\cite{reichenbach-2007-448}.
In the absence of noise, the  equations still predict the formation of spiral waves, but their
spatial arrangement depends on the initial conditions.

Support of the German Excellence Initiative via the program ``Nanosystems
Initiative Munich" is gratefully acknowledged.
M.~M. is grateful to the Humboldt Foundation
for support through the grant IV-SCZ/1119205. \\
{\small *Currently at: Mathematics Institute \& Centre for Complexity Science,
University of Warwick, Coventry CV4 7AL, U.K.}

\newpage

\begin{widetext}

\begin{center}
{\Large {\bf \textsf{ Supplementary Material}} \textsf{(EPAPS Document)}\\
{\small \textsf{http://netserver.aip.org/cgi-bin/epaps?ID=E-PRLTAO-99-046747}}}
\end{center}

\section{From the master equation to stochastic partial differential equations}

In this Supplementary Notes, we want to explicitly outline how, starting from 
 the master equation associated with the stochastic May-Leonard model [defined by the reactions~(1)],
the set of stochastic partial differential equations (3) can be obtained via system size expansion.

For the sake of illustration, here we focus on the role of internal noise stemming from the reactions~(1), and ignore spatial degrees of freedom. As detailed in a forthcoming publication~[18], and following the reasoning presented in~[16],  
in a proper continuum limit, the spatial dispersal of individuals is accounted by diffusive terms in the SPDE (3).\\

As in the main text, the overall number of individuals is denoted $N$ and ${\bm s}=(a,b,c)$ stands for the frequencies (or densities) of the species $A$, $B$, and $C$ in the population (i.e. $s_1=a, s_2=b$ and $s_3=c$). 
The master equation giving the time-evolution of the probability $P({\bm s},t)$ of finding the system in the state ${\bm s}$ at time $t$ then reads
\begin{align}
\partial_t P({\bm s},t)=\sum_{\delta{\bm s}}\big\{P({\bm s}+\delta{\bm s},t)\mathcal{W}({\bm s}+\delta{\bm s}\rightarrow {\bm s})
- P({\bm s},t)\mathcal{W}({\bm s}\rightarrow {\bm s}+\delta{\bm s}) \big\}\,,
\label{master_eq}  
\end{align}
where $\mathcal{W}({\bm s}\rightarrow {\bm s}+\delta{\bm s})$ denotes the transition probability from state ${\bm s}$ to the state ${\bm s}+\delta{\bm s}$ within one time step (loss term), $\mathcal{W}({\bm s}+\delta{\bm s}\rightarrow {\bm s})$ is the analogous gain term, and the summation extends over all possible changes $\delta{\bm s}$. As an example, the relevant changes $\delta s_1\equiv \delta a$ in the density $a$ resulting from the basic reactions (1) are $\delta s_1=\delta a=-1/N$ in the third reaction,  $\delta s_1=1/N$ in the fourth, and zero in all others.
We also choose the unit of time such
that, on average, every individual reacts once per time step. The transition rates resulting from the reactions~(1) then read  $\mathcal{W}=N\sigma ac$ for the reaction 
$CA\stackrel{\sigma}{\longrightarrow} C\oslash$
(the prefactor of $N$ enters due to our choice of  time scale, where $N$ reactions occur in one unit of time) and $\mathcal{W}=N\mu a(1-a-b-c)$ for $A\oslash\stackrel{\mu}{\longrightarrow} AA$. Transition probabilities associated with all other reactions (1) follow similarly.

The system size, or Kramers-Moyal, expansion (SZE)~[16] of the Master equation is an expansion in the increment $\delta{\bm s}$, which is proportional to $N^{-1}$. Therefore, the SZE may be understood as an expansion in the inverse system size $N^{-1}$. To second order in $\delta{\bm s}$, it yields  the (generic) Fokker-Planck equation [16]:
\begin{equation}
\partial_tP({\bm s},t)=-\partial_i[\alpha_i({\bm s})P({\bm s},t)]+\frac{1}{2}\partial_i\partial_j[\mathcal{B}_{ij}({\bm s})P({\bm s},t)] ~.
\label{fokker_planck}
\end{equation}
For the system under consideration,  in the above the indices $i,j\in \{1,2,3\}$ 
and  the summation convention in (\ref{fokker_planck}) implies sums carried over them. According to the  Kramers-Moyal expansion (or SZE), the quantities 
$\alpha_i$ and $\mathcal{B}_{ij}$ are given by [16]
\begin{align}
\alpha_i({\bm s})=&\sum_{\delta{\bm s}} \delta s_i\mathcal{W}({\bm s}\rightarrow {\bm s}+\delta {\bm s})\,,  \cr
\mathcal{B}_{ij}({\bm s})=&\sum_{\delta {\bm s}}\delta s_i \delta s_j\mathcal{W}({\bm s}\rightarrow {\bm s}+\delta{\bm s}) \,.
\end{align}
Note that $\mathcal{B}$  is symmetric.
For the sake of clarity, we outline the calculation of $\alpha_1({\bm s})$: The relevant changes $\delta s_1=\delta a$ result from the third and fourth reactions in~(1), as described above. The corresponding rates respectively read  $\mathcal{W}=N\sigma ac$ and  $\mathcal{W}=N\mu a(1-a-b-c)$, resulting in $\alpha_1({\bm s})=\mu a(1-a-b-c)-\sigma ac$.
The other quantities are computed analogously. 
All explicit expressions for $\alpha_i({\bm s})$ and  $\mathcal{B}_{ij}({\bm s})$ will be derived and given in detail
in [18]. The well-known correspondence between Fokker-Planck equations and Ito calculus [16] implies that~(\ref{fokker_planck}) is equivalent to the following set of Ito stochastic differential equations
(with the above  summation convention):
\begin{align}
\partial_t a&=\alpha_1 + \mathcal{C}_{1j}\xi_j\,,\cr
\partial_t b&=\alpha_2 + \mathcal{C}_{2j}\xi_j\,,\cr
\partial_t c&=\alpha_3  + \mathcal{C}_{3j}\xi_j \,,
\label{stoch_part_eq}
\end{align}
where the matrix $\mathcal{C}$ is defined from $\mathcal{B}$
via the relation $\mathcal{C}\mathcal{C}^T=\mathcal{B}$ [16], and the $\xi_i$'s denote (uncorrelated) Gaussian white noise terms. Note that for the model under consideration $\mathcal{B}$ is diagonal 
 and one can therefore always choose  a diagonal matrix for $\mathcal{C}$ [see Eq.~(5)], with only $\mathcal{C}_{ii}$'s contributing to the right-hand side of Eqs.~(11).

In Ref.[18], we demonstrate that (in a proper continuum limit) spatial degrees of freedom and exchange processes 
simply yield additional diffusive terms $\nabla^2 s_i$ in (\ref{stoch_part_eq}). This leads to the SPDE (3), given and discussed 
in the main text, in which the $\xi_i'$s still denote Gaussian white noise contributions.

\section{Generalization in arbitrary dimensions}

The SPDE~(3) and the CGLE~(6), as well as the analytical predictions~(7), are valid in all spatial dimensions. While in the text, for the
sake of specificity, we mainly focus on the two-dimensional situation, 
here we comment on some  features of the one-dimensional
version of the system, as well as on some  properties of the general
situation in higher dimensions.

In one dimension and in the absence of exchange processes (mixing), our model is expected to exhibit coarsening like the
cyclic Lotka-Volterra model (see e.g.~[12]). The same phenomenon still occurs in the presence of very slow mixing rate
(for the model under consideration, this means very low values of the exchange rate $\epsilon$).
On the other hand, and as shown in the main text, in the presence of (finite) mixing the systems' behavior is aptly described by the SPDE~(3). The underlying CGLE~(6) predicts the propagation of traveling waves, with velocity and wavelength still given by the analytical expression~(7).  Similarly to what has been found in the two-dimensional system~[6], if the exchange rate is ``moderate'', i.e. below a certain mobility threshold but still finite [6,15], the system confirms these predictions and the species coexist. However, due to stochastic events and the presence of absorbing boundaries, some domains will occasionally merge, resulting in growing domains. This coarsening phenomenon will happen on a much longer time-scale than in the absence of the mixing processes.  Above the threshold value for the diffusivity, the system is  well-mixed, the underlying spatial structure plays no role,  and the description in terms of the rate equations (2) is valid. In this mean-field scenario, the system approaches an absorbing steady
state and no patterns emerge.

In higher dimensions (i.e. in dimensions $d\geq2$), the description of the stochastic spatial system in terms of the SPDE~(3) and of the CGLE~(7) is again valid for ``moderate'' (or intermediate) mixing (i.e. for a finite mobility rate which is below a certain critical threshold, see [6]). In this situation, and as discussed in the main text, the description in terms of the CGLE~(6) is qualitatively valid and predicts the emergence of moving spiral waves in two dimensions (which is the case discussed in detail in the main text, see also [6,15]), and to 
``scroll waves'', i.e. vortex filaments, in three dimensions [24].

\newpage
\end{widetext}

%\bibliographystyle{apsrev}

%\bibliography{lit_rps}

\end{document}